\documentclass[aps,superscriptaddress,floatfix,twocolumn]{revtex4-1}
\pdfoutput=1
\usepackage{amsmath,amssymb,eucal,graphicx,color}

\begin{document}

\title{Spectral content of a single non-Brownian trajectory}

\author{Diego Krapf}
\affiliation{Department of Electrical and Computer Engineering, Colorado State
University, Fort Collins, CO 80523, USA}
\affiliation{School of Biomedical Engineering, Colorado State University, Fort
Collins, CO 80523, USA}
\author{Nils Lukat}
\affiliation{Institute for Materials Science, Biocompatible Nanomaterials,
University of Kiel, Kaiserstrasse 2, 24143 Kiel, Germany}
\author{Enzo Marinari}
\affiliation{Dipartimento di Fisica, Sapienza Universit{\`a} di Roma, P.le A.
Moro 2, I-00185 Roma, Italy}
\affiliation{INFN, Sezione di Roma 1 and Nanotech-CNR, UOS di Roma, P.le A.
Moro 2, I-00185 Roma, Italy}
\author{Ralf Metzler}
\affiliation{Institute for Physics and Astronomy, University of Potsdam,
Karl-Liebknecht-Str 24/25, 14476 Potsdam-Golm, Germany}
\author{Gleb Oshanin$^*$}
\affiliation{Sorbonne Universit\'e, CNRS, Laboratoire de Physique Th\'{e}orique
de la Mati\`{e}re Condens\'{e}e (UMR 7600), 4 Place Jussieu, 75252 Paris Cedex
05, France}
\email{oshanin@lptmc.jussieu.fr}
\author{Christine Selhuber-Unkel}
\affiliation{Institute for Materials Science, Biocompatible Nanomaterials,
University of Kiel, Kaiserstrasse 2, 24143 Kiel, Germany}
\author{Alessio Squarcini}
\affiliation{Max-Planck-Institut f\"ur Intelligente Systeme, Heisenbergstr. 3,
70569, Stuttgart, Germany}
\affiliation{IV. Institut f\"ur Theoretische Physik, Universit\"at Stuttgart,
Pfaffenwaldring 57, 70569 Stuttgart, Germany}
\author{Lorenz Stadler}
\affiliation{Experimental Physics I, University of Bayreuth, D-95440 Bayreuth,
Germany}
\author{Matthias Weiss}
\affiliation{Experimental Physics I, University of Bayreuth, D-95440 Bayreuth,
Germany}
\author{Xinran Xu}
\affiliation{Department of Electrical and Computer Engineering, Colorado State
University, Fort Collins, CO 80523, USA}

\begin{abstract}
Time-dependent processes are often analysed using the power
spectral density (PSD), calculated by taking an appropriate Fourier transform
of individual trajectories and finding 
the associated ensemble-average.
Frequently, the available experimental data sets are too small for such ensemble
averages, and hence it is of a great conceptual and practical importance
to understand to which extent relevant information can be gained from
$S(f,T)$, the PSD of a single trajectory. Here we focus on the behavior of
this random, realization-dependent variable, parametrized by frequency $f$
and observation-time $T$, for a broad family of anomalous diffusions---fractional
Brownian motion (fBm) with Hurst-index $H$---and derive exactly its
probability density function. We show that $S(f,T)$ is proportional---up to
a random numerical factor whose universal distribution we determine---to
the ensemble-averaged PSD. For \textit{subdiffusion\/} ($H<1/2$) we find
that $S(f,T)\sim A/f^{2H+1}$ with random-amplitude $A$. In sharp contrast,
for \textit{superdiffusion\/} $(H>1/2)$  $S(f,T)\sim BT^{2H-1}/f^2$ with
random amplitude $B$. Remarkably, for $H>1/2$ the PSD exhibits the same
frequency-dependence as Brownian motion, a deceptive property that may lead to
false conclusions when interpreting experimental data. Notably, for $H>1/2$
the PSD is \textit{ageing\/} and is dependent on $T$. 
Our predictions for both  sub- and superdiffusion are confirmed
by experiments in live cells and in agarose hydrogels, and by extensive
simulations.
\end{abstract}

%\pacs{87.80.Nj; 05.40.Jc; 05.40.-a; 02.50.Cw}
%{\rm Key Words:} Anomalous diffusion, fractional Brownian motion, power spectral density, probability distribution
%

\maketitle 

\section{Introduction}
 
The power spectral density of any time-dependent process $X_t$ is a fundamental
feature of its spectral content, dynamical behavior and temporal correlations
\cite{norton}.  It is an important measure for various processes across many
disciplines, including loudness of musical recording \cite{voss,geisel}, evolution of climate data
\cite{talkner}, time gaps between large earthquakes \cite{sornette}, retention times of chemical tracers in
groundwater \cite{kirchner},
noise in graphene devices \cite{balandin},  fluorescence intermittency in nanosystems \cite{fran},
 current fluctuations
in nanoscale electrodes \cite{krapf}, stochastic processes with random reset \cite{satya},  some extremal properties of Brownian motion \cite{krapivsky},
diffusion in strongly disordered systems \cite{enzo,dean},
and ionic currents across nanopores \cite{ramin1},
to name a few diverse examples.

In its standard  definition, the power spectral density (PSD) is the Fourier transform of the
autocorrelation function of $X_t$ over an infinitely large observation time
$T$, i.e., it is an ensemble-averaged property taken in the limit $T\to\infty
$. In many situations, however, one cannot create a sufficiently large
statistical sample to achieve a reliable ensemble average,  and even though
the limit $T \to \infty$ can be formally taken in mathematical expressions,
it cannot be reached in experiments. Instead, one often deals with either a single
or  a few individual finite-length realizations of the process, particularly,
in experimental data dealing with {\it in vivo} systems \cite{noerregaard},
climate change \cite{climatechange}, or financial markets \cite{eco}. In
this regard, a question of immense conceptual and practical importance is
whether one can learn relevant information about the system from the PSD of
just a single or a few finite-length realizations.

Several recent studies examined the PSD from such single-trajectory data. 
Power spectra of individual time-series were examined for a
stochastic model describing blinking quantum dots \cite{nie,QD} and
also for single-particle tracking experiments with tracers in artificially
crowded fluids \cite{mat}. Notably, the scaling exponent of the power
spectrum computed from a few single-trajectory PSDs remains very stable.
In addition, the power spectra of the velocity of  motile amoeba revealed
a robust large-$f$ asymptotic behavior of the form $1/f^{2}$, for all
measured individual trajectories \cite{li,ped}.

Without a solid mathematical theory, these observations 
\cite{nie,QD,mat,li,ped}  may be considered merely as curious
coincidences. However, in a recent work \cite{diego}  (see also the perspective \cite{ulrich}), it was proven that
for standard Brownian motion, the single-trajectory PSD $S(f,T)$ in the large-$f$
limit and at finite $T$ exhibits the same $f$-dependence as its
traditional ensemble-average counterpart. This mathematical prediction was
fully corroborated by numerical simulations and experiments with polystyrene
beads in aqueous solution \cite{diego}. 

Despite its ubiquitous
appearance in nature \cite{frey}, Brownian motion is just a particular example
of a stochastic process, and there is no evidence that the same behavior
should hold for other naturally occurring transport processes. In this regard,
it seems highly desirable to have an analogous proof for anomalous diffusion,
with mean-squared displacement (MSD)
\begin{equation} 
\label{MSD}
\langle X_t^2 \rangle \sim t^{\alpha},
\end{equation}
and anomalous diffusion exponent $\alpha \neq 1$, where the brackets here and henceforth denote the
average over the statistical ensemble. 

Such processes are  widely observed  in soft matter, condensed matter and biological
systems, e.g., diffusion in viscoelastic and crowded systems,
the motion of proteins \cite{noerregaard,rienzi} or sub-micron tracers in living
cells \cite{lene,tabei}, in artificially crowded liquids \cite{lene1,szymanski},
telomere diffusion in the cell nucleus \cite{garini}, diffusion in disordered
media \cite{disorder}, dynamics of ultra-cold atoms \cite{atmos} and in
lipid membranes \cite{prx,bba,sadegh,membranes}. Anomalous diffusion is also found in
other systems, including heartbeat intervals \cite{heartbeat}, DNA sequence
landscapes \cite{dna}, and even in the daily fluctuations of climate variables
\cite{climate} and economic markets \cite{eco}.

Here we  calculate {\em exactly}
for any $T$,
$f$ and $\alpha$
 the full probability density function (PDF) of a single-trajectory PSD $S(f,T)$ - a random,  realization-dependent variable -  for the widely observed
process of  fractional Brownian motion (fBm) \cite{ness,
pccp}  (see also Sec. \ref{fbm} for more details). 
Analogous to the parental fBm process, the PDF of the PSD of its individual realizations appears to be entirely characterized by its two first moments: 
we thus derive an explicit expression  for the ensemble-averaged PSD 
\begin{align}
\label{mean}
\mu = \mu(f,T) = \langle S(f,T)\rangle  \,,
\end{align}
the first moment of the PDF,  which is a standard property, and also go beyond the textbook definition and 
determine its variance 
\begin{align}
\sigma^2 = \sigma^2(f,T) =  \langle S^2(f,T)\rangle  -  \langle S(f,T)\rangle^2 \,.
\label{var}
\end{align}
This
permits us to quantify
the effective broadness 
of the PDF via its coefficient of variation $\gamma = \sigma/\mu$. 
We realize that, (for any $f$ and $T$ and regardless of the 
value of the anomalous diffusion exponent),
  $\gamma$ always exceeds the value $1$ such that the standard deviation $\sigma$ of the single-trajectory PSD is always greater than its ensemble-averaged value $\mu$. This implies that the PDF is broad and cannot be characterized exhaustively solely by its first moment,  which  justifies {\em a posteriori} our quest for the form of the full PDF.  Moreover, we find that the value achieved by $\gamma$ in the limit $f T \to \infty$ is very meaningful, and on this basis we
  offer a novel and very robust criterion, which will  
  permit to prove the anomalous character of random motion in situations when the analysis of the MSD deduced from experimental data leads to ambiguous conclusions.
  
 Our theoretical analysis then culminates at the observation that for sufficiently large values of $f$ a single-trajectory PSD $S(f,T)$ is linearly proportional (with a universal, dimensionless, random proportionality factor) to its mean value $\mu$, which embodies the full dependence on $T$ and $f$.  This generalizes the previous observation made in Ref. \cite{diego} for standard Brownian motion to a wide class of anomalous diffusion.  
 Here, however, the value of the anomalous diffusion exponent appears to be crucially important: for $\alpha < 1$ (subdiffusion) the PSD attains a stationary form $1/f^{\alpha+1}$ for sufficiently large $f$ and $T$, while in the superdiffusive case ($\alpha > 1$)   the leading behavior of the PSD is given by  
 $T^{\alpha - 1}/f^2$, i.e., the PSD
 is {\em ageing} and is deceivingly proportional to $1/f^2$, 
 where the exponent $2$ characterizing the $f$-dependence is the same as for standard Brownian motion ($\alpha = 1$) \cite{diego}, regardless of the actual value of $\alpha > 1$. In consequence, one should exercise a  great deal of care in the interpretation of the data for superdiffusive motion and rather concentrate on the ageing behavior on $T$ than on the $f$-dependence.  

Further on, we compare a variety of our analytical predictions 
against the corresponding  analysis of
single-trajectory data garnered from  experiments in quite
diverse systems: the 
dynamics of telomeres in the nucleus of live cells,  of polystyrene microspheres in agarose hydrogels, of motile
{\em Acanthamoeba castellanii} and their intracellular vacuoles.  
Some features of the predicted behavior of the PDF, which we could not access in experiments, are also verified
by 
an extensive numerical analysis. As we will show, our analytical predictions on the $f$-dependence of the PSD in the 
subdiffusive case, ageing behavior and the deceptive $1/f^2$ dependence in the superdiffusive case, as well as the corresponding distributions 
of the universal random amplitude
are fully in
line with experimental observations in biologically relevant systems.

We remark that the four systems used in our experimental analysis are just a few particular examples of systems with an fBm-type dynamics. In general, fBm encompasses a broad range of naturally occurring
processes with continuous paths and long-ranged temporal correlations, which entail
both sub- and superdiffusive behaviors, depending on whether the increments of particle
displacement are negatively or positively correlated. In particular, physically fBm
processes describe the over-damped, antipersistent motion of particles in
viscoelastic environments \cite{goychuk,mat,pccp}, as well as the persistent
superdiffusion in actively driven systems \cite{christine}. 
The characteristic antipersistent signature of subdiffusive fBM 
was identified in the dynamics of chromosomal loci and RNA-protein particles in live
bacterial cells \cite{weber}, lipid granules in yeast cells in the millisecond range
\cite{lene}, tracer beads in worm-like micellar solutions
\cite{lene1}, lipid molecules in dilute bilayer membranes in
supercomputing studies \cite{prx,prl,kneller}, and chromatin in
Langevin dynamics simulations \cite{dipierro}. With different observables
fBm-type motion was further identified in the dynamics of chromosomal telomeres in living
U2OS cells \cite{burnecki} and nanosized particles in crowded dextran solutions
\cite{szymanski,ernst,mat}. 

FBm is thus a very generic
stochastic process, and it combines both subdiffusive and superdiffusive motion in a
common framework, and therefore we regard it here as the first prototype example for the study
of single-trajectory PSD.  
Of course, fBm does not cover all possible kinds of anomalous diffusion
\cite{pccp}. Moreover, in some instances, fBm dominates the dynamics of a system at
intermediate time scales and it is tempered to become standard diffusion at longer times;
or dynamical transitions between different types of fBm may take place \cite{igor,
daniel}. A systematic analysis of other representative examples of anomalous diffusion,
of combinations of different anomalous diffusions, and of processes with dynamical
transitions between different types of behavior is thus ultimately necessary, in order
to attain a full understanding  of the spectral content of a single-trajectory PSD.
In turn, such an analysis will provide robust criteria  permitting eventually 
to distinguish between different types of random motion. Our work thus represents an
essential first step in this direction.

The outline of this paper is as follows: In Sec. \ref{fbm} we describe the statistical properties of fractional Brownian motion, present the definitions of the random variables of interest here, namely, the power spectral density of individual trajectories of fBm in case of one-dimensional dynamics, as well as for a more general case of a $d$-dimensional dynamics with projections on the coordinate axes.  We also present the definition of the moment-generating function of the PSD of individual fBm trajectories. The 
desired PDF follows from the latter upon
a mere inversion of the Laplace transform. In Sec. \ref{Sec_03} we describe our experimental systems which exhibit anomalous, non-Brownian dynamics and also briefly recall how both the MSD and the PSD can be deduced 
from the experimental data. At the end of this Section we also describe the algorithm of our numerical analysis. Sec. \ref{Sec_04} presents our main exact analytical results and a discussion of their asymptotic behavior. On this basis, we formulate here a robust, novel criterion which will permit to prove the anomalous or normal (standard Brownian) character of dynamics. 
This criterion is based on a statistical sample and is validated by numerical simulations.   Further on, in Sec. \ref{Sec_06} we compare our analytical predictions against the results of simulations and experimental data garnered from experiments performed for four different systems exhibiting an anomalous behavior. In Sec. \ref{Sec_05} we present a brief summary of our results and a perspective.

\section{Fractional Brownian motion and its power spectral densities}
\label{fbm}

FBm is a Gaussian stochastic process  and hence, is entirely characterized by its first moment and the covariance, which defines its auto-correlation at two different time instants $t_1$ and $t_2$. FBm has
zero mean value and  its covariance function is given by 
\begin{equation}
\label{covarFBM}
\langle X_{t_1} X_{t_2}\rangle=D\left[t_1^{2H}+t_2^{2H}-|t_1-t_2|^{2H}\right] \,,
\end{equation}
where $H \in (0,1)$ is the so-called Hurst index \cite{ness,pccp}. Comparing the expression in Eq. \eqref{covarFBM} for $t_1 = t_2$ with Eq. \eqref{MSD}, one infers that for fBm 
the
anomalous diffusion exponent $\alpha$ is simply related to $H$, $\alpha = 2 H$. 

Standard Brownian motion,  
on which the analysis in Ref. \cite{diego} was concentrated
 is recovered for a particular case $H=1/2$ only. In this case, 
 the increments of the process are independent and $D$  is the standard
diffusion coefficient. When $H>1/2$ (corresponding to $\alpha > 1$), the increments
are positively  correlated such that if there is an increasing pattern in
the previous steps, it is likely that the current step will be increasing
as well, resulting ultimately in a superdiffusive motion. 
For $H<1/2$ the
increments are negatively correlated, such that it is most likely that after an increasing step a decreasing one will follow. This ultimately entails a
 subdiffusive motion. In
the two latter cases $D$ can be thought of as  a  proportionality factor
with units ${\rm length}^2/{\rm time}^{2H}$.

We focus here on
the single-component single-trajectory PSD 
\begin{equation}
\label{1spec}
S(f,T)=\frac{1}{T} \left|\int^T_0 \exp(ift)X_tdt\right|^2, 
\end{equation}
which is a random variable dependent
on a given realization $X_t$ of a one-dimensional fBm and is parametrized by
the observation time $T$ and the frequency $f$. For its generalization
over the $d$-dimensional case, we represent a trajectory ${\bf R}_t$ of a
$d$-dimensional fBm  as ${\bf R}_t=\{X_t^{(1)},X_t^{(2)},\ldots,X_t^{(d)}\}$
\cite{1}. Here $X_t^{(j)}$ is the projection of ${\bf R}_t$ onto the axis $x_j$
and is statistically independent of other components. With this definition we
consider the $k$-component version of a $d$-dimensional single-trajectory PSD,
\begin{equation}
\label{2spec}
S_k(f,T) = \frac{1}{T} \sum_{j=1}^{k} \left|\int^T_0 \exp\left(i f t\right) X_t^{(j)}dt\right|^2,
\end{equation}    
where  $k=1,\ldots,d$, is the number of the tracked components. For $k=1$
Eq.~\eqref{2spec} reduces to $S(f,T)$.

We note that the standard text-book definition of the PSD is based on the ensemble-averaged expressions in Eqs. \eqref{1spec} and \eqref{2spec}.
Our aim here is much more ambitious: we proceed to calculate exactly the full
PDF of the random variable $S_k(f,T)$ for arbitrary
$H$, $k$, $f$ and $T$. This can be done rather straightforwardly, 
if we manage to determine the moment-generating function of the random variable $S_k(f,T)$, defined formally
by
\begin{equation}
\label{MGF}
\Phi_{\lambda} =  \left \langle \exp\left( -  \frac{\lambda}{T} \sum_{j=1}^{k} \left|\int^T_0 \exp\left(i f t\right) X_t^{(j)}dt\right|^2\right) \right \rangle \,,
\end{equation}
with $\lambda \geq 0$.  The desired PDF $P(S_k)$ follows from Eq. \eqref{MGF} by a mere inversion of the Laplace transform with respect to the parameter $\lambda$. Exact results for both the moment-generating function and the PDF, as well as 
asymptotic expressions of the first two moments of the PDF  are presented in Sec. \ref{Sec_04}. Details of the derivations, which are rather lengthy, and also quite cumbersome exact expressions for the first two moments of the PDF (valid for arbitrary values of the parameters $f$ and $T$, and for arbitrary $H \in (0,1)$)
are presented in the Supplemental Material (SM).

\section{Experimental and numerical analyses}
\label{Sec_03}

\subsection{Experimental systems}

We study experimentally the dynamics of polystyrene microspheres in agarose hydrogels,
of telomeres in the nucleus of live cells, of motile
{\em Acanthamoeba castellanii}  amoeba and of their intracellular vacuoles.  

\subsubsection*{Agarose hydrogel}
We recorded the motion of 50-nm microspheres in agarose hydrogel. A 1.5\% agarose gel was prepared from agarose powder (Cat. 20-102GP, Genesee Scientific, San Diego, CA) without further purification by dissolving it in phosphate-buffered saline. Carboxylate-modified polystyrene microspheres with 50-nm nominal diameter (Cat. PC02002, Bangs Laboratories, Fishers, IN) were first heated to 60$^\circ$C in 0.5\% Tween20 and introduced into the agarose solution also at 60$^\circ$C. The agarose/microsphere solution was allowed to mix at 60$^\circ$C for 15 min and then transferred to a hot glass-bottom petri dish and left to slowly cool to room temperature.

The microspheres were imaged in an inverted microscope equipped with a 40x objective (Olympus PlanApo, N.A. 0.95) and a sCMOS camera (Andor Zyla 4.2) operated at 71 frames per second. The first 2,048 images were used for further tracking and analysis. Tracking of the microspheres in the plane was performed in LabView using a cross-correlation based tracking algorithm \cite{tracking}. Immobile particles and particles that exhibited very little motion were discarded. A total of 20 trajectories were analysed in terms of their PSD.   

\subsubsection*{Telomeres}
Trajectories of telomeres in the nucleus of untreated U2OS 
cells were acquired at 8 frames per second and evaluated as described before \cite{SW2017}. 
It was shown previously that the time-averaged mean squared displacement 
(TA-MSD) of these trajectories featured an fBm-like sub-diffusive scaling 
for short and intermediate times with a mean exponent 
$\langle\alpha\rangle\approx0.5$. From these previously analysed data, 
we selected $19$ individual trajectories, each of 2,500 frames length, with scaling exponents 
in the range $\alpha=0.5\pm 0.05$ for PSD analysis. 

\subsubsection*{Amoeba and intracellular vacuoles}
Trajectories of amoeba and intracellular vacuoles were recorded using {\em A. castellanii} cultured as previously described \cite{christine}. Imaging was done using  a Hamamatsu ORCA ER2 camera on an Olympus IX71 microscope and images were recorded with the MATLAB Image Acquisition Toolbox (Mathworks, Inc.) at 9 frames per second. In addition, every two seconds the image was segmented using an edge detection algorithm in MATLAB and the centre of mass of the amoeba was calculated. To record over long time periods, the amoeba was kept in the centre of the image by automatically moving along a scanning stage (M{\"a}rzh{\"a}user, SCAN IM 112 x 74). In addition, the position of intracellular vacuoles was detected using a home-written segmentation algorithm in MATLAB. In brief, first edge detection was carried out, followed by a Hough transformation to find circles and an algorithm to verify the vacuoles by their light edge. The centre of the circles in the images was determined and gave the position of the vacuoles. All trajectories were optically verified. Previously,  it was shown that the vacuole intracellular motion within {\em A. castellanii} is super-diffusive \cite{christine}.

\begin{figure*}
\includegraphics[width=18cm]{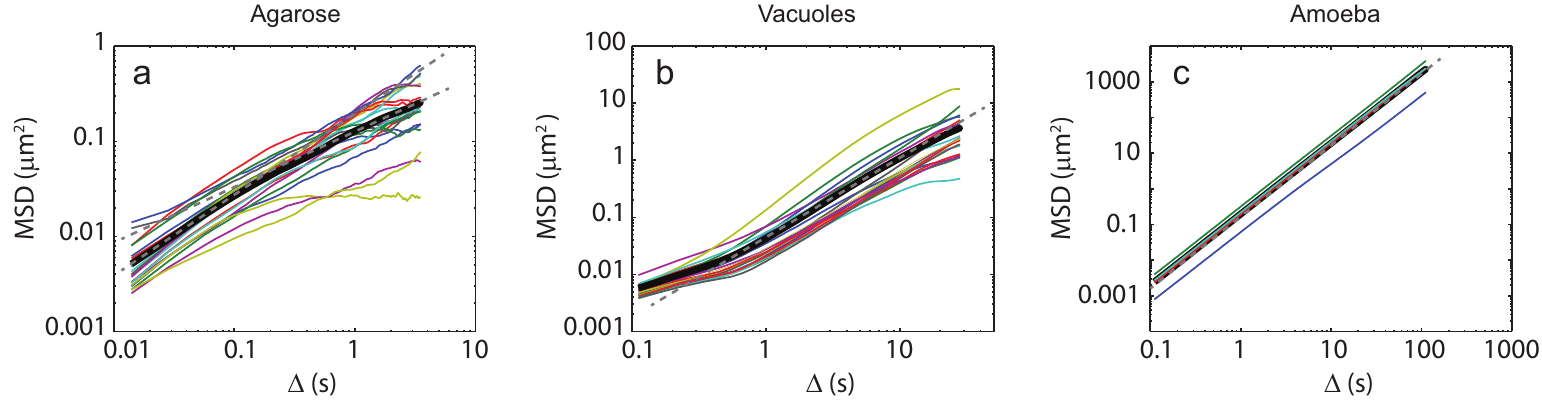}
%\begin{figure}[ht]
%\begin{center}
%\centerline{\includegraphics[width = 1 \textwidth]{MSD.eps}}
\caption{MSD analysis. (a) Time-averaged MSD (TA-MSD), Eq. \eqref{TAMSD} as a function of lag time $\Delta$ for 20 individual microsphere trajectories together with their ensemble average. (b) TA-MSD for 20 individual representative vacuole trajectories together with the ensemble average from 50 trajectories. (c) TA-MSD for four individual amoeba trajectories together with their ensemble average. In all three panels the thick black lines show the ensemble averages and the grey dashed lines show power law behaviour MSD $\sim \Delta^\alpha$.   
\label{MSDF}
}
%\end{center}
\end{figure*}

Four individual amoeba trajectories, each consisting of 16,384 frames, were analysed in terms of their PSD. Short vacuole trajectories were discarded and 50 vacuoles trajectories, all from the same cell, were analysed. In order to avoid differences in trajectory lengths, only the first 2,048 frames in each vacuole trajectory were used in the analysis.  

\subsection*{Mean squared displacement (MSD)}
The time-averaged MSD of individual trajectories $X_p$, 
where $p$ is an appropriately discretized time variable, is
defined as (see, e.g., Ref. \cite{pccp})
\begin{equation}
\label{TAMSD}
\overline{\delta^2(\Delta)}=\frac{1}{N-\Delta} \sum_{p=1}^{N-\Delta}\left(X_{p+
\Delta}-X_p \right)^2 .
\end{equation}
This property was computed in MATLAB as a function of the lag time $\Delta$ for all analysed trajectories. The MSD of representative microspheres, amoeba, and vacuole trajectories along with their ensemble mean are shown in Fig. \ref{MSDF}. The MSD from the analysed telomere trajectories were previously reported \cite{SW2017}. The anomalous exponent as obtained from the MSD is for microspheres $\alpha_S=0.87$ for short times and $\alpha_L=0.61$ for longer times (Fig. \ref{MSDF}a); for telomeres $\alpha=0.5$ (see Ref. \cite{SW2017}); for vacuoles $\alpha=1.33$ (Fig. \ref{MSDF}b); and for amoeba $\alpha=1.97$ (Fig. \ref{MSDF}c).

\subsection*{PSD analysis}
Single- and two-component PSDs of individual trajectories (as defined in Eqs. \eqref{1spec} and \eqref{2spec}, respectively), were obtained
 in MATLAB from the Fourier-transformed components $X_t^{(1)}$ and $X_t^{(2)}$ of three-dimensional trajectories. 
Care was taken that all trajectories of the same type included the same number of data points and the same frame rate.

For analysing the fluctuations of the PSD, i.e., to obtain the empirical distributions of the amplitudes of the PSD for sub- and super-diffusive cases, the gross scaling $1/f^{\beta}$ was obtained from the ensemble-averaged PSD, where $\beta=2H+1$ for the sub-diffusive cases (microspheres and telomeres) and $\beta=2$ for the super-diffusive ones (vacuoles). From these data sets we extracted values $A' = S(f,T)f^{\beta}$ in the following frequency ranges: (i) $11 \mathrm{ Hz} < f <87\mathrm{ Hz}$ for microspheres, (ii) $1 \mathrm{ Hz}<f<10\mathrm{ Hz}$ for telomeres, and (iii) $1 \mathrm{ Hz} <f< 5 \mathrm{ Hz}$ for vacuoles. We did not extract the fluctuations of the amoeba because only four trajectories were used. Then, we normalised the fluctuations according to $A=A'/\langle A'\rangle$. The same procedure was followed to obtain $B$ for the vacuoles. These data were then compared to the theoretical predictions as described  in Sec. \ref{Sec_04} below.
 
\subsection*{Numerical algorithms}

Numerical simulations of fBm are far more
complicated than the ones used, for example, for a standard Brownian
motion. FBm is not
a Markov process and 
has long range correlations.  In order to reproduce fBm numerically we use the
exact Davies Harte Circulant method (see, e.g., Refs. \cite{DH,WC,DN,D,MD}). Due to the use
of Fast Fourier Transform the required CPU time for reproducing a
$T$ steps trajectory is of order $T \log(T)$ (and not of order $T^2$
as a naive approach would give). The Davies-Harte approach is a very
powerful exact method, and for samples of the size we use its running
time is comparable to the one of effective approximate methods \cite{D}. We
use trajectories of $T=2^{21}$ to $T=2^{23}$ discrete time steps.
The total CPU time we have used for all the numerical runs that have
been useful to prepare this work is of the order of few months of one
core of Intel(R) Xeon(R) CPU E5-2620 0 $@$ 2.00GHz.

\section{Analytical predictions}
\label{Sec_04}

Our first step consists in calculating
the moment-generating function $\Phi_{\lambda}$ of the $k$-component single-trajectory PSD (see Eq. \eqref{2spec}), 
defined in Eq. \eqref{MGF}. We obtain (see SM for the details of the derivation)
\begin{align}
\label{Phi}
\Phi_{\lambda}=\left[1+2\mu\lambda+\left(2-\gamma^2\right)\mu^2\lambda^2\right]^{
-k/2},
\end{align}
where $\mu$ is the first moment of
a single-component single-trajectory PSD, Eq. \eqref{mean}, $\sigma^2$ is the variance of this random variable, Eq. \eqref{var},  and
$\gamma=\sigma/\mu$ is the coefficient of variation of the PDF of a single-component single-trajectory PSD $S(f,T)$, Eq. \eqref{1spec}. 
Inverting the Laplace
transform with respect to $\lambda$ we readily obtain the PDF of $S_k(f,T)$,
\begin{eqnarray}
\label{dist}
&&P\left(S_{k}(f,T)=S\right) = \frac{\sqrt{\pi}}{2^{\frac{k-1}{2}} 
\Gamma\left(k/2\right) \sqrt{2 - \gamma^2} \left(\gamma^2 -1 \right)^{\frac{k-1}{4}}}  \nonumber\\
&\times& \frac{S^{\frac{k-1}{2}}}{ \mu^{\frac{k+1}{2}}} \exp\left(- \frac{1}{2 - \gamma^2} \frac{S}{\mu}\right) {\rm I}_{\frac{k-1}{2}}\left(\frac{\sqrt{\gamma^2 - 1}}{2 - \gamma^2} \frac{S}{\mu}\right) \,,
\end{eqnarray} 
where ${\rm I}_{\nu}(z)$ is the modified Bessel function of the $1$st kind. We emphasise that the expressions in Eqs. \eqref{Phi} and \eqref{dist} are exact and hold for any $f$, $T$ and also for any value of the Hurst index $H$. We note that $\Phi_{\lambda}$ and $P\left(S_{k}(f,T)=S\right) $ are entirely defined by the first
two moments of $S(f,T)$---and hence, all higher moments of the $k$-component single-trajectory PSD $S_{k}(f,T)$ can be
expressed solely through the first two moments of $S(f,T)$. This is a direct consequence
of the Gaussian nature of the parental process $X_t$.  This suggests, in turn, 
that the expressions in Eqs. \eqref{Phi} and \eqref{dist} may hold, in general, for arbitrary Gaussian processes, not necessarily for the fBm only. The dependence $\mu$, $\sigma^2$ and hence, of $\gamma$ on the characteristic parameters 
will depend, of course, on the case at hand.

For fBm processes, the exact dependence
of $\mu$, $\sigma$ and, hence, of $\gamma$ on $f$ and $T$ for any value of $H \in (0,1)$ is presented
in SM.  Below we discuss their rather complex 
behavior focusing first on the coefficient of variation $\gamma$, which characterizes the
effective broadness of the PDF in Eq. \eqref{dist}.

\begin{figure}
\includegraphics[width=8.6cm]{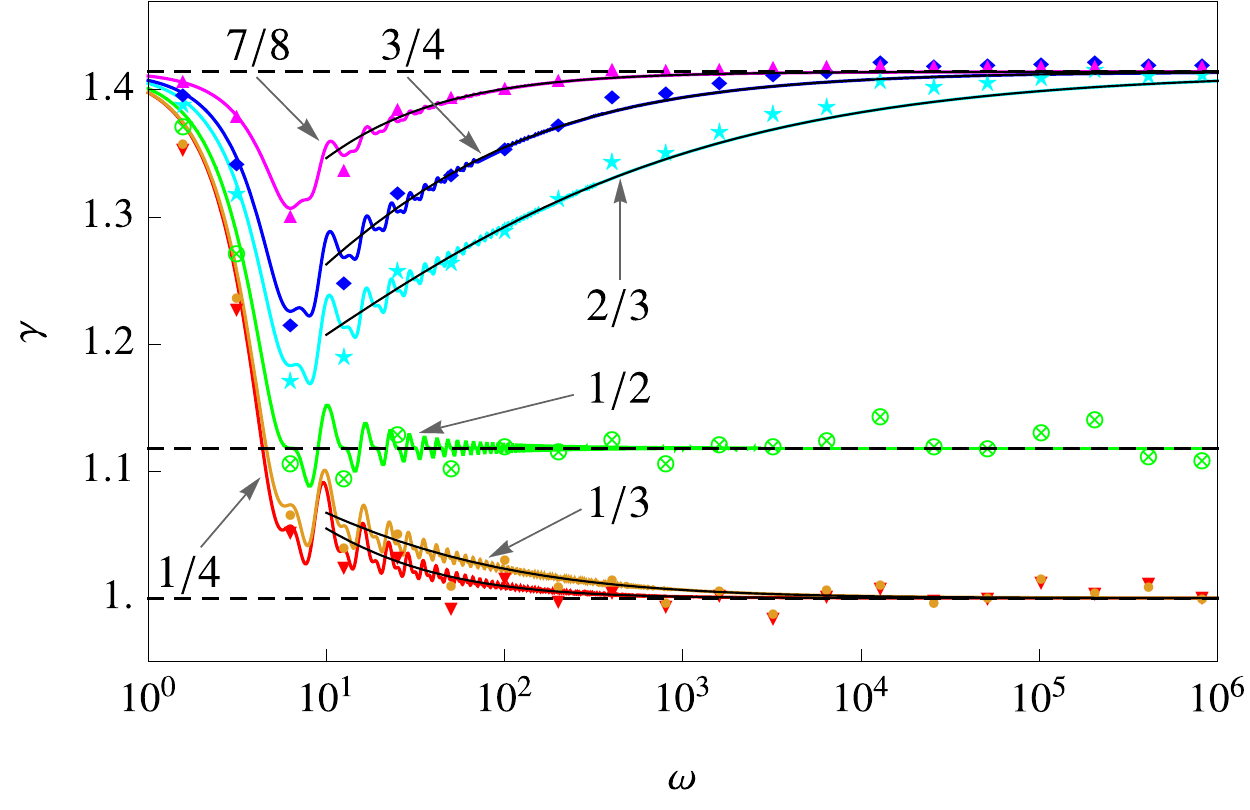}
\caption{Coefficient of variation $\gamma$ as a function of
$\omega=f T$. Colored solid curves represent exact values of $\gamma$ (arrows
indicate the corresponding values of $H$),  defined by Eq. (47) in the SM, while thin solid curves depict the
asymptotic expression in Eq. \eqref{gammas}. Horizontal dashed lines correspond to
$\sqrt{2}$ (top), $\sqrt{5}/2$ (middle) and $1$ (bottom). Symbols represent numerical results averaged over $10^4$
realizations of trajectories consisting of $T=2^{23}$ discrete time steps.}
\label{FIG1}
\end{figure}

The coefficient of variation $\gamma$,  which enters Eqs.~\eqref{Phi} and \eqref{dist}, is a
dimensionless numerical factor that depends on $f$ and $T$ only through the
function $\omega=fT$. Figure~\ref{FIG1} shows $\gamma$ as a function of $\omega$ for six different Hurst indices spanning the range $1/4 \le H \le
7/8$. The behaviour of $\gamma$ has several characteristic features, which
can be clearly observed in Fig. \ref{FIG1}: \\
(i) In the limit $\omega\to 0$, the
coefficient $\gamma$ tends to the universal value $\sqrt{2}$, regardless of
the value of $H$. Next, $\gamma$ is an oscillatory function of $\omega$,
and the oscillations are prominent at moderate values of $\omega$. In the
limit $\omega\gg1$ the oscillatory terms fade out and $\gamma$ is given by
very simple asymptotic formula (see SM for derivation)
\begin{align}
\label{gammas}
\gamma\sim\left[1+\left(1+c_H\omega^{1-2 H}\right)^{-2}\right]^{1/2}, 
\end{align}
with $c_H=\Gamma\left(1+2H\right) \sin\left(\pi H\right)$. This asymptotic form is depicted by thin solid curves in Fig.~\ref{FIG1}.
\\
(ii) We see that $\gamma \geq 1$, in the whole range of variation of $\omega$. This signifies that the standard deviation of the single-component single-trajectory PSD always exceeds its mean value. In consequence, the PDF in Eq. \eqref{dist} is effectively broad and the analysis of the power spectrum using the standard ensemble-averaged PSD $\mu$ only is rather meaningless.\\
(iii) A most remarkable
feature - rendering $\gamma$ a crucial and highly practical property
for fBm-type processes  - is that it offers the sought criterion for
anomalous diffusion, since the values attained by $\gamma$ in the limit
$\omega\to\infty$ are distinctly different: $\sqrt{2}$, $\sqrt{5}/2$, and $1$,
independent of the exact value of $H$ but solely dependent on whether one has
a superdiffusive ($H>1/2$), diffusive ($H=1/2$), or subdiffusive ($H<1/2$)
behaviour, respectively. 
These analytical predictions 
are fully confirmed by numerical simulations for a number of $H$ values.

Before we proceed, it may be expedient to dwell some more on the last point.   When dealing with particle-tracking experiments, one often observes
values of $\alpha$ that are only slightly different from $1$. Consequently,
in these cases, it is not obvious whether one is dealing with anomalous diffusion, or simply if
the fitting of the curves started too early and includes transient behavior. On the other hand,
the asymptotic value of $\gamma$ at large frequencies provides, in principle, an immediate answer to this
question and reveals whether the underlying diffusion process is normal or anomalous. 
Such an unequivocal confirmation of anomalous diffusion can provide 
extremely valuable evidence to 
drive efforts into searching for microscopic mechanisms underlying
the dynamics and lead eventually to a deeper comprehension of the processes in the system under study.  

Note that here, however, we resorted to proof-of-concept numerical simulations, because the confirmation of this prediction
requires a rather big statistical sample, which we were unable to create in current experimental analysis. We nonetheless perform such an analysis below in Sec. \ref{Sec_06} (see Fig. \ref{FIG10}): it appears to be instructive and shows that even a small sample containing only few tens of trajectories can provide a meaningful representation of the overall trend. Given the current rapid progress in single particle tracking techniques, sufficiently large experimental samples are certainly within reach.

Even though $\gamma$ allows finding whether the process is subdiffusive or superdiffusive, 
it does not permit one to deduce the value of the anomalous  diffusion exponent
$\alpha$. Below we discuss how one can find this further piece of the puzzle by analyzing the asymptotic behavior of the ensemble-averaged PSD $\mu$ and the corresponding limiting behavior of the PDF in Eq. \eqref{dist}. Consider first the case of subdiffusion ($H<1/2$). We suppose that $\omega$
is sufficiently large, such that $\gamma-1\leq\varepsilon$, where $\varepsilon$
is a small parameter. In virtue of relation \eqref{gammas} the above inequality
holds for $\omega$ within the interval $\omega\in(\omega_l^{(\mathrm{sub})},
\infty)$ where $\omega_l^{(\mathrm{sub})}=1/(2c_H\varepsilon)^{1/(1-2H)}$
(e.g., for $H=1/4$ and $\varepsilon=0.01$ one gets $\omega_l\approx6.4\times
10^3$). In this limit, the denominator in Eq.~\eqref{Phi} becomes a full
square, i.e., $\Phi_{\lambda} \simeq \left[ 1+\mu\lambda \right]^{-k}$, with
accuracy set by $\varepsilon$. This means, in turn, that the PDF of $S_k(f,T)$ becomes, up to terms of order of
$\varepsilon$, the gamma distribution with shape parameter $k$ and scale
parameter $\mu$. Consequently, in this case, the $k$-component single-trajectory PSD obeys
the equality in distribution
\begin{equation}
\label{a}
\frac{S_k(f,T)}{\mu^{\mathrm{(sub)}}}\overset{d}{=} A +O(\varepsilon),
\end{equation}
where the omitted terms are small in $\varepsilon$ and $A$ is a random
numerical factor with distribution
\begin{equation}
\label{A}
P(A)=A^{k-1}\exp(-A)/\Gamma(k).
\end{equation}

In the superdiffusive case ($H>1/2$), we again assume
that $\omega$ is sufficiently large such that the inequality
$\sqrt{2}-\gamma\leq\varepsilon$ holds. By virtue of relation \eqref{gammas},
this is true when $\omega\in(\omega_l^{( \mathrm{sup})}, \infty)$ with
$\omega_l^{(\mathrm{sup})}=(2\sqrt{2}c_H/\varepsilon) ^{1/(2H-1)}$ (e.g.,
for $H=3/4$ and $\varepsilon = 0.01$, we have that $\omega_l^{(\mathrm{sup})}
\approx 7.1\times10^4$, i.e., a somewhat bigger value than the one in the
subdiffusive case). In this limit the coefficient in front of the term
quadratic in $\lambda$ in the denominator in Eq.~\eqref{Phi}, (i.e., $(2 - \gamma^2)$), is less
than $\varepsilon$ such that $\Phi_{\lambda} \simeq \left[ 1+2\mu\lambda
\right]^{-k/2}$ and, in turn, the PDF of $S_k(f,T)$
becomes the gamma distribution with scale $2\mu$ and shape parameter
$k/2$. Consequently, the $k$-component single-trajectory PSD follows the equality in
distribution
\begin{equation}
\label{b}
\frac{S_k(f,T)}{\mu^{\mathrm{(sup)}}}\overset{d}{=} 2 B+O(\varepsilon),
\end{equation}
where $B$ is a random numerical factor with distribution
\begin{equation}
\label{B}
P(B)=B^{k/2-1}\exp(-B)/\Gamma(k/2).
\end{equation}
Therefore, the equalities in Eqs.~\eqref{a} and \eqref{b} suggest that for
both the subdiffusive and superdiffusive cases  the single-trajectory PSD should always be linearly proportional
to its ensemble-average value $\mu$, (which incorporates the dependence
on frequency),  at large values
of $\omega$. The proportionality factor is merely a random number with
distribution given by Eqs. \eqref{A} or \eqref{B}, which does not entail any
additional dependence on $f$ or $T$.

Below we specify the spectral content of $\mu$. In the SM we show that for
subdiffusive fBm at sufficiently large values of $T$ and $f$, $\mu$ has the
scaling form
\begin{equation}
\label{limi1}
\mu^{\mathrm{(sub)}} =\frac{2c_HD}{f^{2H+1}}.
\end{equation}
In the superdiffusive case $H > 1/2$, at large $T$ and $f$, 
\begin{equation}
\label{asymp2}
\mu^{\mathrm{(sup)}} =\frac{2D}{f^2}T^{2H-1}+\frac{2c_HD}{f^{2H+1}}+o\left(1\right),
\end{equation}
where the Landau symbol $o(1)$ states that the omitted terms vanish as
$T\to \infty$. Result \eqref{asymp2} unveils two remarkable features of
the  ensemble-averaged PSD in the superdiffusive case:\\
(i) First, regardless
of the value of $H$, for large $T$, the frequency dependence has the
universal $1/f^2$ form, precisely that of the PSD for standard Brownian
motion. Therefore, experimental analyses of the frequency dependence of the
PSD in the superdiffusive case may lead to the false conclusion that one
deals with Brownian motion ($H=1/2$). Consequently, one should exercise care
in interpreting data in this case: while Brownian motion has a PSD that
scales as $1/f^2$, the observation of exclusively such a dependence does
not guarantee that one indeed deals with Brownian motion. \\
(ii) Second, a crucial
difference from Brownian motion is the dependence of the amplitude  on the
observation time $T$. This {\it ageing\/} behavior can be used to distinguish
the $T$-independent PSD for Brownian motion from the superdiffusive case:
$H$ can be deduced by analysing the spectrum at some fixed frequency given
that one expects $S_k(f,T)\sim T^{2H-1}$.

Lastly, we show that 
the value of $H$ can be deduced from the
spectrum
evaluated at zero frequency, 
\begin{equation}
S_k(f=0,T) = \frac{1}{T}  \sum_{j=1}^{k} \left(\int^T_0 X_t^{(j)}dt\right)^2 \,,
\end{equation}
which represents the sum (divided by $T$) of squared areas under the projections of the random curve $X_t$ on different axes.
In the SM we show that
 the ensemble-averaged PSD
at zero frequency is universally (for both subdiffusive and superdiffusive $H$) described by
\begin{equation}
\label{0}
 \mu(f=0,T) = \frac{D T^{2H+1}}{(H+1)} \,,
 \end{equation}
 (see also the result in Ref. \cite{satya} with the reset rate set equal to zero). On the other hand, the variance of the single-trajectory PSD obeys (see SM), again for any $H$,
 \begin{equation}
 \label{00}
 \sigma^2 (f=0,T) = \frac{2 D^2 T^{4H+2}}{(H+1)^2} \,.
 \end{equation}
 This signifies that the coefficient of variation $\gamma = \sqrt{2}$, regardless of the value of $H$, such that the PDF of $S_k(f=0,T)$ is 
 the gamma distribution with scale $2 \mu(f=0,T)$ and shape parameter $k/2$ for any $T$.  In consequence,  $S_k(f=0,T)$ obeys
exactly  the single-trajectory relation in Eq. \eqref{b} with the correction term $O(\varepsilon)$ identically equal to zero implying
that the Hurst index can be deduced directly from the PSD at zero frequency. 
 
 Below we explore both possibilities to deduce $H$ from a single-trajectory data, taking advantage of the ageing behavior of $S_k(f,T)$ and of the dependence of the PSD at zero frequency on the observation time $T$.

\section{Comparison with experimental and numerical data}
\label{Sec_06}

\begin{figure*}
\includegraphics[width=18.3cm]{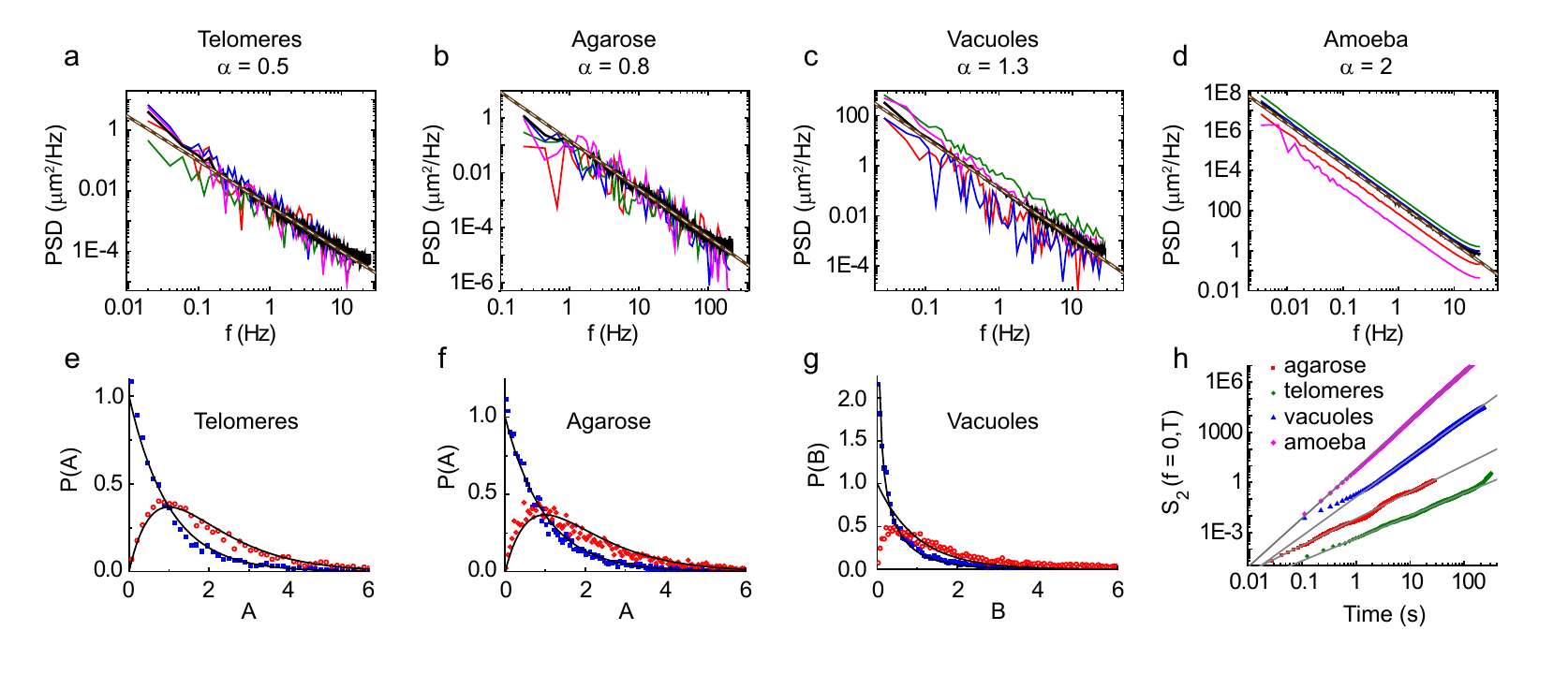}
\caption{Power spectrum analysis of experimental data sets.  (a-d)
PSD of representative trajectories along with the ensemble-averaged PSD
for telomeres in the nucleus of HeLa cells, 50-nm microspheres in 1.5\%
agarose gel, intracellular vacuoles within amoeba, and the motion of amoeba.
The dashed thick lines show the $1/f^{1.49}$ trend for panel (a), $1/f^{1.76}$
for panel (b) and $1/f^2$ for panels (c) and (d). In each case, the PSDs of
four trajectories are presented (log-sampled with a factor 1.1 for clarity)
together with the ensemble-averaged PSD (thicker black lines, $n=19, 20, 50,$
and $4$ trajectories for telomeres, microspheres, vacuoles, and amoeba,
respectively). (e-g) Distribution of amplitudes of the PSD for one and two
components. (h) PSD evaluated at zero frequency. The zero frequency spectra
are shifted for clarity and displayed together with the fitted power-law
functions (grey solid lines). The experimental results throughout agree
excellently with the theoretical predictions.}
\label{FIG2}
\end{figure*}

We tested our predictions for the PSD of single trajectories in four different
experimental data sets and multiple numerical simulations. The experimental
data consist of two systems exhibiting subdiffusive behavior and two
systems exhibiting superdiffusion.  For subdiffusive dynamics,
we analysed the motion of 50-nm microspheres in agarose gels and telomeres
in the nucleus of mammalian cells \cite{SW2017}. For superdiffusive
behaviour, we studied the motion of live amoeba and their intracellular
vacuoles. Representative MSD of individual trajectories in all these systems
are presented in Fig.  \ref{MSDF} along with their respective averages
of the time-averaged MSDs. Examples of PSDs of the single trajectories are
shown in Figs. \ref{FIG2}a-d.

The time-averaged MSDs of telomeres scale with an exponent $\alpha=0.5$,
(i.e., $H=0.25$) for short and intermediate times \cite{SW2017}, predicting a
PSD $S(f,T) \sim A/f^{1.5}$. As shown in Fig. \ref{FIG2}a,  the individual
trajectories agree with this prediction and the ensemble-averaged PSD from 19
trajectories yields  $\mu^{(\mathrm{sub})}\sim 1/f^{1.49}$. 
We also show that the experimentally observed fluctuations in
the PSDs remarkably confirm the predicted universal distribution Eq. \eqref{A}  for both
one- and two-components PSDs, i.e., for $k=1$ and $k=2$, respectively
(Fig. \ref{FIG2}e).  Similar agreements are found for the motion of
$50$-nm  microspheres in $1.5\%$ agarose gel. As shown in Fig. 
\ref{MSDF}a,
the MSD of these particles scales with an exponent $\alpha=0.87$ ($H=0.43$)
for short times and $\alpha=0.61$ ($H=0.30$) for long times. The PSD yields
$\mu^{(\mathrm{sub})}\sim 1/f^{1.76}$ (Fig. \ref{FIG2}b), and the PSD
fluctuations also follow closely a gamma distribution (Fig. \ref{FIG2}f)
as predicted by Eq.~\eqref{A}.

The motion of amoebae and their intracellular vacuoles are good examples
of superdiffusive dynamics. Intracellular vacuoles are subdiffusive at
short lag times and superdiffusive with $\alpha=1.33$ at long lag times
(Fig.  \ref{MSDF}b). This behaviour is typical of active motion in the
cytoplasm \cite{ActiveMotion}. Interestingly, the MSDs of the centre of mass
of the investigated amoebae show almost ballistic motion with $\alpha=1.97$
(Fig.  \ref{MSDF}c). The PSDs of the motion of both the amoebae  and the
vacuoles therein,  clearly show the predicted deceptive $1/f^2$ behavior
(Figs. \ref{FIG2}c and d). The distribution of the PSD amplitudes is also
shown for the vacuoles in Fig. \ref{FIG2}g together with the predicted
gamma distributions, Eq.~\eqref{B}, revealing an excellent agreement with
the latter for $k=1$. The discrepancy with our two-component analytical
prediction for small $B$-values is likely associated with small amplitude
antipersistent motion of the vacuoles,  as is evident from 
 the trajectories.

In Fig. \ref{FIG2}h we present the averaged spectra
at zero frequency for both subdiffusive and superdiffusive cases ( see eqs. \eqref{0} and \eqref{00} in Sec. \ref{Sec_04}). We used $19$, $20$, $50$ and $4$ trajectories for the telomeres, microspheres, vacuoles
and amoeba, respectively, to get directly the Hurst exponents: $H=0.30$, 
$0.18$, $0.67$ and $1.92$. Despite the small sizes of our statistical samples,
the obtained values of $H$ agree well with the values deduced from
the corresponding MSDs. 

\begin{figure}
\includegraphics[width=8.6cm]{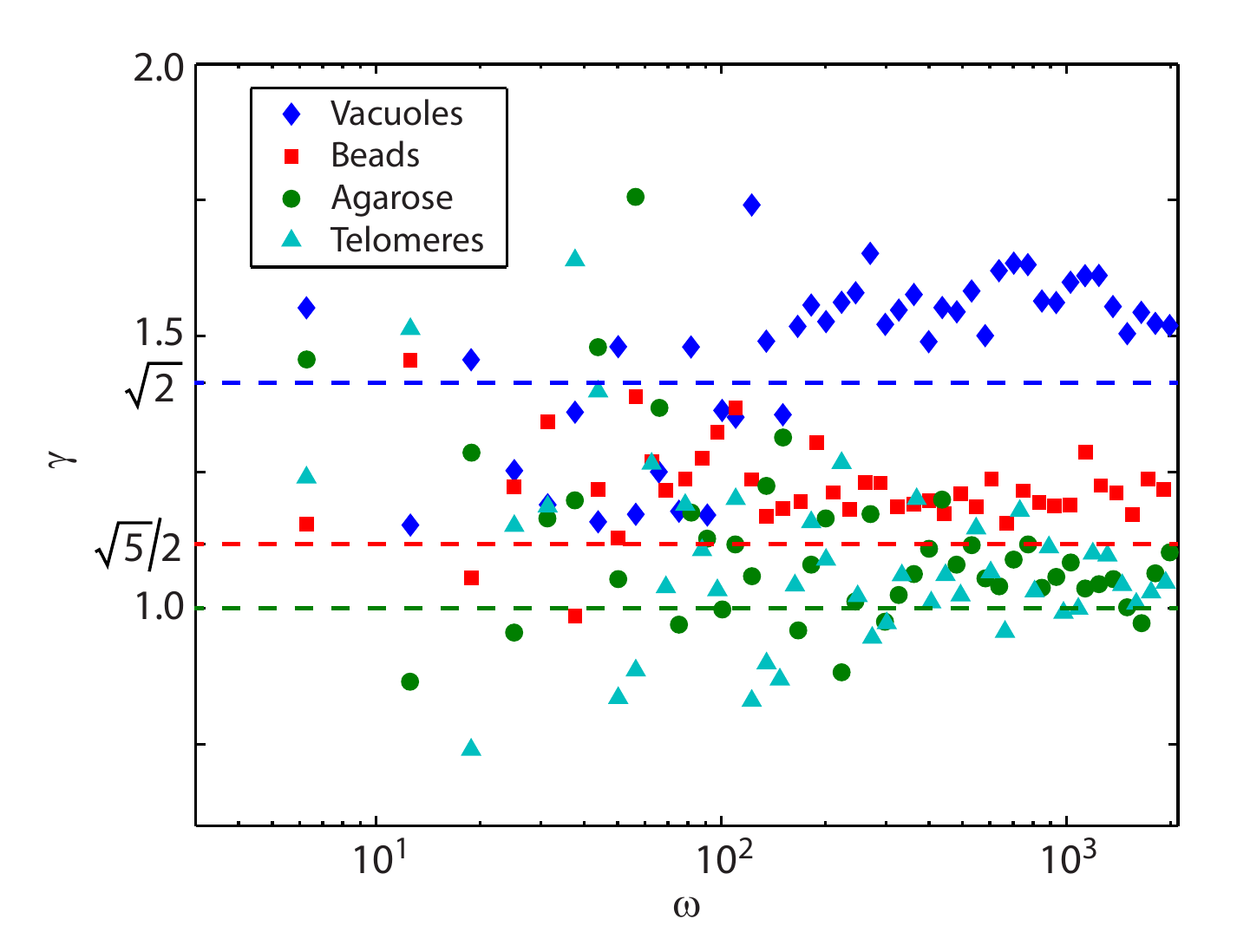}
\caption{Coefficient of variation $\gamma$ as a function of
$\omega=fT$ in small statistical samples of experimental data. Dashed horizontal lines highlight the analytical predictions made
in Sec.~\ref{Sec_04} with blue, red, and green indicating the values
$\gamma=\sqrt{2}$ (superdiffusive fBm), $\gamma=\sqrt{5}/2$
(standard Brownian motion \cite{diego}), and $\gamma= 1$
(subdiffusive fBm), respectively.   
Symbols represent the values of $\gamma$ drawn from experiments. Blue diamonds: vacuoles ($50$ trajectories); red squares: $1.2$-$\mu$m beads in aqueous solution ($150$ trajectories) \cite{diego}; green circles: $50$-nm microspheres in agarose hydrogel ($20$ trajectories); and cyan triangles: telomeres in the nucleus of mammalian cells ($19$ trajectories).
}
\label{FIG10}
\end{figure}

We revisit next the behavior of the coefficient of variation $\gamma$ of the single-trajectory PDF (see Fig. \ref{FIG1}) and address the question whether meaningful information
can already be drawn from small statistical samples of experimental data.   In Fig. \ref{FIG10} we plot the value of $\gamma$ as a function of $\omega = f T$ obtained from only $19$ experimentally recorded trajectories of telomeres, $20$ trajectories of microspheres in agarose hydrogels and $50$ intracellular vacuole trajectories, as well as from a larger number of trajectories ($150$) of micrometer-sized beads in an aqueous solution \cite{diego}.  The microspheres in aqueous solution provide an excellent example of standard Brownian motion, i.e., $H=0.5$. One observes 
that, indeed,  in the large-$\omega$ limit,  $\gamma$ converges 
to distinctly different values for superdiffusion, normal diffusion and subdiffusion cases.  
For vacuoles, at large $\omega$, the coefficient of variation $\gamma$ is observed to converge to $1.55\pm0.01$, for Brownian motion (beads in aqueous solution), it is observed to converge to $1.21\pm0.01$, for telomeres to $1.05\pm0.01$ and for the microspheres in agarose gels to $1.07\pm0.01$. In line with our analytical prediction of a universal value of
$\gamma$ for subdiffusive fBm, the obtained values for telomeres
and microspheres are very close to each other. Overall, the
experimentally determined values for $\gamma$ are only about 10\%
larger than our analytical predictions ($\sqrt{2} \approx 1.41$
for vacuoles, $\sqrt{5}/2 \approx 1.12$ for beads in aqueous
solution \cite{diego}, and $1$ for telomeres and microspheres).
Given the small size of the statistical sample, we consider such
a favorable agreement quite remarkable. In comparison, the perfect
agreement of our predictions with $\gamma$ values from fBm simulations
(cf. Fig.~\ref{FIG1}) rather represents an exceptional situation due
to the big statistical sample ($10^4$ trajectories). Moreover, in
experiments many different, sometimes uncontrollable factors, e.g.
detector noise, may come into play which leads to an increasing
variance of the single-trajectory PSD and hence to elevated values
of $\gamma$. We plan to examine this important aspect in more detail
in our future work.

\begin{figure*}
\includegraphics[width=19cm]{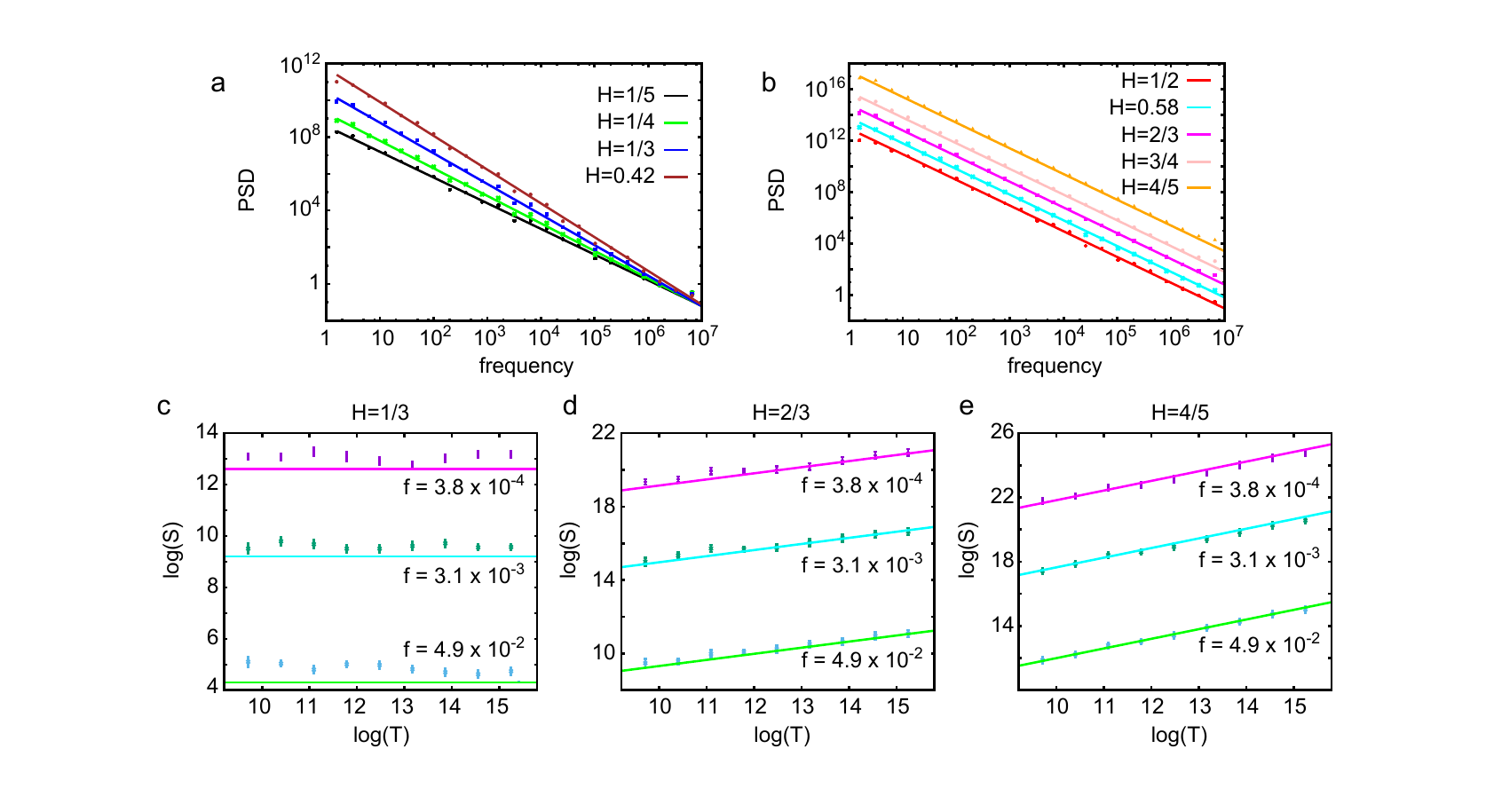}
\caption{Numerical confirmation of our theoretical predictions. (a)
Frequency dependence of a single-trajectory PSD in the subdiffusive case
(Eqs.~\eqref{a} and \eqref{limi1}) for $H = 0.42, 1/3, 1/4$, and $H = 1/5$,
averaged over $10$ realisations. Straight lines have slope $-(2H + 1)$, as
shown in Eq.~\eqref{limi1}. The symbols represent the results of numerical
simulations. (b) Deceptive $1/f^2$-dependence (Eq.~\eqref{asymp2}) in the
superdiffusive case. Symbols represent results of numerical simulations
for individual trajectories averaged over $10$ realisations. (c) Apparent
independence of a single-trajectory PSD in the subdiffusive case on the
observation time for sufficiently large $T$. (d-e) Ageing behavior of
a single-trajectory PSD in the superdiffusive case for three different
values of the frequency. Straight lines have slopes $2H-1$ as shown in
Eq.~\eqref{asymp2}. Symbols represent numerical results averaged over
$50$ realisations.}
\label{FIG3}
\end{figure*}

We further performed extensive analyses of single-trajectory PSDs for
different values of $H$ using numerical simulations. Figures \ref{FIG3}a and
b show the results for single-trajectory PSDs as a
function of the frequency (for sufficiently large $T$) for different values
of $H$ between $1/5$ and $4/5$. Namely, the subdiffusive cases ($H<1/2$)
are shown in Fig. \ref{FIG3}a and other cases ($H > 1/2$) are shown in
Fig. \ref{FIG3}b.  One observes excellent agreement between the predicted
behaviour, Eqs. \eqref{a} and \eqref{limi1}, and the numerics even for a small
statistical sample consisting of $10$ realisations. In Fig. \ref{FIG3}c we
also demonstrate that the single-trajectory PSD for a specific subdiffusive
case ($H=1/3$) is not ageing.  On the other hand, Figs. \ref{FIG3}e and
f illustrate the ageing behavior of a single-trajectory PSD for $H =2/3$
and $4/5$, at three fixed frequencies. Here, the straight lines indicate the
predicted ageing dependence $T^{2 H - 1}$, Eq. \eqref{asymp2}, while the
symbols represent the results of numerical simulations averaged over $50$
realisations. We again observe a perfect agreement with our theoretical
predictions.

\section{Discussion}
\label{Sec_05}

In summary, we here combined theoretical, numerical and experimental analyses
to provide a comprehensive answer to the conceptually and practically important
question: which information can be reliably obtained from the spectral content
of a single realisation of naturally occurring anomalous-diffusion processes.
Given the widespread occurrence of $1/f$-type of spectra in the analysis of
experimental systems and signals across almost every field of physics, 
such an analysis is very pertinent.
Focusing on a wide class of such processes---the so-called fractional
Brownian motion---we derived exactly the distribution of a single-trajectory
power-spectral density (PSD) and analyzed its asymptotic forms for both subdiffusive 
and superdiffusive dynamics. On this basis, we unveiled several striking
features: \\
(i) At a fixed observation time and in the limit of high frequencies,
this distribution reduces to simple forms with a unique scaling given by
the ensemble-averaged PSD, which incorporates the full dependence on $f$
and $T$. As a consequence, one expects that for an arbitrary realization
of the process a single-trajectory PSD should exhibit the same large-$f$
dependence as a traditional ensemble-averaged PSD.\\
 (ii) Our experiments
and numerical simulations impressively evidence that this is indeed the
case for both super- and subdiffusive fBm-type processes. For subdiffusive processes,
the exponent characterising the spectrum is equal to $2 H+1$ and hence,
the anomalous diffusion exponent can be obtained by evaluating the slope of
the PSD. For superdiffusive processes, in contrast, the exponent is deceptively universal
and equal to two, which can lead to the false conclusion that one deals with
ordinary Brownian motion, while in reality the process is superdiffusive. 
We find this prediction particularly important since it will permit to avoid a misinterpretation of experimental results.\\
(iii) For superdiffusive processes the amplitude of the PSD is ageing, i.e.,
dependent on the observation time. However, it is difficult to observe this
dependence on a single trajectory since the $T$-dependence is weaker than the
large fluctuations between nearby frequencies.  Here, a statistical sample (comprising, however, only $50$ trajectories)
was used in order to observe the ageing trend and to extract the value of
the anomalous diffusion exponent from the ageing behavior. \\
(iv) We
showed that the coefficient of variation $\gamma$ of a single-trajectory PSD provides
a novel criterion for anomalous diffusion. For fBm, its large-$f$ form  assumes
only three different values, depending on whether we observe subdiffusion,
normal diffusion, or superdiffusion. Our analytical predictions are in a perfect agreement with the results
of numerical simulations for a representative statistical sample ($10^4$ trajectories), but are also in line 
with the experimental results, obtained from a fairly small statistical sample ($19$ to $50$ trajectories). \\
(v) Lastly, our theoretical, numerical and experimental analysis shows unequivocally 
that the coefficient of variation always exceeds the value $1$, meaning that the standard deviation of a single-trajectory PSD 
is generically bigger than its mean value. 
In standard nomenclature of the statistical analysis, the distributions which possess such a property are considered to be effectively broad. 
This implies that the analysis of the spectral content of individual trajectories in terms of only the ensemble-averaged PSD has limited meaning,
which justifies completely our quest for the full PDF of this important characteristic property.

To conclude, from an experimental perspective,
our results serve as a reliable framework in the interpretation of noisy
data obtained from a single trajectory 
- it has become routine to garner few individual particle trajectories of impressive length
 in the wake of superresolution microscopy and super-
computing. In perspective, our results will thus play an important role in
extracting more physical information from them.

Finally, we remark that especially in the complex environment of biological
cells, where a vast array of specific and non-specific interactions transpire,
fBm does not account, of course, for all possible types of observed anomalous
diffusions. Therefore, additional stochastic mechanisms may be superimposed,
such as short-time or even simultaneous scale-free trapping time dynamics
\cite{lene,tabei,weigel,weron17}.   In other instances, fBm may
be tempered or there may occur dynamical transitions between different types
of fBm \cite{igor,daniel}. Extensions of our analysis over other possible kinds of
anomalous diffusion, such as the fBm models with dynamical transitions
\cite{igor}, ``diffusing diffusivity'' models \cite{chechkin}, scaled Brownian motion
\cite{sbm}, or continuous-time random walks with a broad distribution of waiting times
\cite{scher} are necessary in order to get a full understanding of
the behavior of the PSD in experimentally relevant systems. 
We believe that our work presents an important first step towards such an
understanding and will prompt a systematic case-by-case analysis.

\acknowledgments

We thank Eli Barkai for discussions. 
D.K. acknowledges the support of the National Science Foundation
under grant no. 1401432.
Research of E.M. is supported in part by the
European Research Council (ERC) under the
European Unions Horizon 2020 Research and Innovation
Program (grant agreement n. 694925).
R.M. acknowledges the German Research Foundation (DFG) grant ME-1535/7-1
and a Humboldt Polish Honorary Research Fellowship from the Foundation for
Polish Science. 
N. L. and C. S.-U. thank the DFG for funding through the Collaborative Research Centre CRC 1261 {\em Magnetoelectric Sensors: From Composite Materials to Biomagnetic Diagnostics}.  MW acknowledges financial support by the VolkswagenStiftung (Az. 92738).\\

\section*{Authors contributions}

D.K. and X.X. performed experimental single-trajectory 
analyses of anomalous diffusion of microspheres in agarose hydrogels, N.L. and C.S-U. studied dynamics of amoebae and their intracellular vacuoles, while L.S. and M.W. performed particle-tracking experiments with telomeres in the nucleus of  mammalian cells. D.K. and M.W. analysed the power-spectra of individual experimental trajectories.  E.M. performed numerical analysis of power spectral densities of fractional Brownian motion. R.M., G.O. and A.S. performed all analytical calculations. D.K., R.M., E.M., G.O., C.S-U. and M.W. have equally contributed to writing the paper. 

\section*{Conflict of interests}

The authors declare no conflict of interests.

\end{document}